\documentclass[aps,prb,longbibliography,reprint,showpacs,twocolumn,superscriptaddress,floatfix]{revtex4-2}
\usepackage{graphicx}
\usepackage{amssymb}
\usepackage{amsmath}
\usepackage[FIGTOPCAP,nooneline]{subfigure}
\usepackage{overpic}
\usepackage{bbold}
\usepackage{color}
\usepackage{physics}
\usepackage{mathtools}

\begin{document}

\title{Light emission in delta-$T$-driven mesoscopic conductors}

\author{M. Hübler}

\affiliation{Fachbereich Physik, Universit\"at Konstanz, D-78457 Konstanz, Germany}

\author{W. Belzig}

\affiliation{Fachbereich Physik, Universit\"at Konstanz, D-78457 Konstanz, Germany}



\date{\today}

\begin{abstract}
The scattering picture of electron transport in mesoscopic conductors shows that fluctuations of the current reveal additional information on the scattering mechanism not available through the conductance alone. The electronic fluctuations are coupled to the electromagnetic field and a junction at finite bias or temperature will emit radiation. The nonsymmetrized current-current correlators characterize the emission and absorption spectrum. Recent interest is focused on the so-called  delta-$T$ noise, which is the nonequilibrium noise caused by a temperature difference between the terminals. Here, we generalize the notion of delta-$T$ noise to the nonsymmetrized current-current correlator at finite frequencies. We investigate the spectral density for energy-independent scattering and for a resonant level as an example of energy-dependent scattering. We find that a temperature difference $\Delta T$ leads to a partial reduction of the noise for certain frequency ranges. This is a consequence of temperature broadening in combination with a frequency shift of the involved Fermi distributions. In the case of energy-independent scattering, the lowest order is a quadratic $\propto (\Delta T)^2$ correction of the thermallike noise spectrum. For the resonance, an additional contribution to the delta-$T$ noise spectrum arises that is $\propto \Delta T$ to the lowest order. 
\end{abstract}


\maketitle


\section{Introduction}
The electron transport in mesoscopic conductors is investigated using the statistics of the electron current, where the first moment corresponds to the average current and the second moment to the noise \cite{Blanter,Levitov_Lesovik_1993,Reulet}. Unavoidable sources of noise are thermal noise at a finite temperature - the so-called Nyquist-Johnson noise \cite{Johnson,Nyquist} and nonequilibrium shot noise \cite{Schottky}. The former is caused by thermal fluctuations in the occupation number and the latter by the stochastic partitioning of charge carriers. The noise at a tunnel junction can be used for primary thermometry \cite{Spietz_Lehnert_Siddiqi_Schoelkopf_2003}. \\
\\
At finite frequencies, the noise involves current operators taken at different times. In general, these operators do not commute, so the symmetrized correlator is studied as an observable \cite{Blanter,Lesovik}. A detector that distinguishes between the transfer of an energy quanta $\hbar \omega$ from or to the conductor can access the nonsymmetrized correlator \cite{Lesovik,Gavish:1}. Indeed, when the fluctuations interact with an electromagnetic field, the energy transfer rate is connected to the nonsymmetrized noise spectrum. Negative frequencies account for the radiated power when one photon is generated in the radiation field and, vice versa, positive frequencies for the absorbed power when one photon is annihilated. In a thermally occupied radiation field, the measured noise power spectrum is a sum of the nonsymmetrized noise spectra at negative and positive frequencies. The prefactors are determined by the Bose-Einstein distribution and, consequently, by the temperature of the electromagnetic field \cite{Lesovik,Gavish:1}. \\
\\
Shaping a possible ac-excitation can strongly influence the noise properties \cite{LEVITOV_Lee_LESOVIK_1996} and can be interpreted as electron-hole pair excitation on the Fermi sea \cite{Vanevic_Nazarov_Belzig_2007}. A noise reduction due to driving was experimentally observed \cite{Dubois_Jullien_Portier_Roche_Cavanna_Jin_Wegscheider_Roulleau_Glattli_2013,Gabelli_Reulet_2013} and measurements at finite frequency reveal a squeezed nonequilibrium state \cite{Gasse_Lupien_Reulet_2013}. \\
\\
A fundamental nonequilibrium noise due to a temperature difference $\Delta T$ was recently demonstrated by Lumbroso and coworkers \cite{Lumbroso,Sivre_Duprez_Anthore_2019,Larocque} in atomic and molecular junctions. This noise, dubbed delta-$T$ noise, is related to the voltage-driven shot noise and inherits the properties of partition noise \cite{Sukhorukov_Loss_1999}. Using the scattering approach, they obtain an approximation of the noise, which is then decomposed into a thermal and a delta-$T$ component. The thermal component corresponds to thermal noise at the average temperature, and the lowest order delta-$T$ component is similar to the quantum shot noise except for different numerical prefactor and scales with $(\Delta T)^2$ instead of the voltage squared. Another study \cite{Larocque} measured and calculated the noise of a voltage-and temperature-biased metallic tunnel junction. This setup operates at a very low temperature and is not restricted to small relative temperature differences. At the limit, when one terminal is at zero temperature and no voltage is applied, the noise has the form of thermal noise with an additional factor $2 \ln 2$ \cite{Sukhorukov_Loss_1999}. \\ 
\\
In a quantum Hall bar furnished with a quantum point contact, the delta-$T$ noise can serve as an instrument to discriminate between electron and quasiparticle tunneling \cite{Rech,rebora:22}. Tunneling of chiral fractional quantum Hall edge states exhibits a negative delta-$T$ noise, in contrast to a positive contribution in the noninteracting case. A sign inversion, from negative back to positive, may also be forced by changing the transmission of the quantum point contact or applying a voltage. The negative signal is attributed to the scaling dimension of the leading charge tunneling operator \cite{He_Dante_2022,Zhang_Igor_2022}. Their results suggest that the negative sign is a property due to many-body interactions. In comparison, a quantum dot in the $SU(2)$ Kondo region has no negative delta-$T$ noise \cite{Hasegawa}, thus the effect does not occur in this case despite the presence of strong correlations. Further, delta-$T$ noise was employed to study experimentally the heat transport of edge modes \cite{Melcer_Dutta_Spanslatt_2022}. \\ 
\\
An investigation of the relative sizes shows that delta-$T$ noise never exceeds the thermallike noise under the zero-average current condition \cite{Eriksson}. In \cite{Eriksson} they studied a resonant level as an example for energy-dependent scattering. In the limit of a small resonance width, the size of delta-$T$ noise approaches the thermallike noise. Furthermore, they investigated noise of heat transport, which is not subject to a limit like charge noise. More recently, bounds on the spin and heat current noise were investigated \cite{Tesser_EtAl_2023}. \\
\\
In this work, we address the nonsymmetrized finite-frequency noise spectrum of a temperature-biased mesoscopic conductor. 
\begin{figure}
\includegraphics{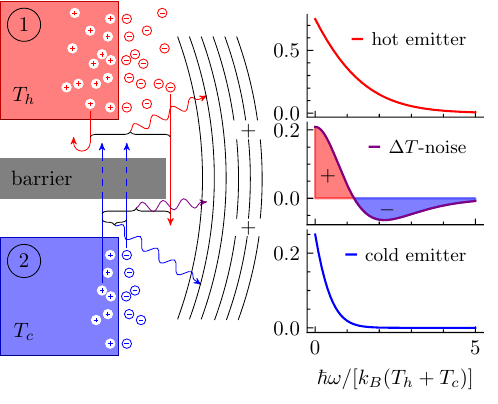}
\caption{The mesoscopic conductor consists of a hot reservoir with a temperature $T_h$ and a cold reservoir with a temperature $T_c$, separated by a potential barrier. In the hot reservoir, more electrons are excited at higher energies, as illustrated by electrons farther away from the surface. The current-current correlator depends on the scattering amplitudes (simplified depicted by the arrows) of electrons (depicted by circles with a minus sign) and holes (depicted by circles with a plus sign) as well as the probability distribution of the terminal from which they originated. Current-current fluctuations are related to the energy transfer rate between the mesoscopic conductor and an electromagnetic field, i.e. to the emission and absorption spectrum. The energy difference $\hbar \omega$ between the electron and the hole corresponds to the energy of the annihilated or excited photon. We separate the spectrum into an equilibrium part (thermallike), a superposition of a hot and cold emitter, and a nonequilibrium part, given by a delta-$T$ noise spectrum. The delta-$T$ noise spectrum is negative at some frequencies, thus diminishing the thermallike noise spectrum. In the shown case, energy-independent scattering is assumed.}
\label{fig:Opener}
\end{figure}
We are interested in separating the light emission and absorption into a thermallike and delta-$T$ spectrum. Figure \ref{fig:Opener} gives an illustrative summary of the considered system and our findings in the case of energy-independent scattering. The mesoscopic conductor is described within the scattering approach, where one terminal assumes a hot temperature and the other a cold temperature. We define the thermallike noise spectrum as the average of the thermal noise spectra at the hot and the cold temperature. Consequently, the delta-$T$ noise spectrum is defined similar to the excess noise spectrum \cite{Blanter,Lesovik_Martin_Torres_1999,Cottet_Doucot_Belzig_2008}. Two distinct contributions to the delta-$T$ noise are obtained, $S^{\Delta T}_1(\omega)$ comes from the correlations of occupied and free electronic states with different Fermi statistics, and $S^{\Delta T}_2(\omega)$ from occupied and free states with the same statistics but associated with a different rate per recombination event than assumed in the thermallike noise. If the scattering is energy independent or the same from both sides, the latter contribution vanishes. Our main result is that the delta-$T$ noise spectrum can get negative at some frequencies, reducing the thermallike noise spectrum. Below, we investigate the spectra for energy-independent scattering and a single resonant level model. In the resonant case  $S^{\Delta T}_1(\omega)$ has a suppressed negative part for the chosen parameters and additionally shows a contribution $S^{\Delta T}_2(\omega)$.\\
\\
The work is structured as follows. In the second section \ref{sec:Scattering}, we introduce the scattering approach and summarize the connection to light emission and absorption. Afterwards, we define the thermallike noise spectrum and discuss the delta-$T$ noise spectrum for general scattering matrices. As an application, we consider energy-independent scattering in section \ref{sec:IndepEnergy} and then a resonant level in section \ref{sec:Resonance}. Our results are recapped in section \ref{sec:conclusions}.

\section{Scattering Approach to delta-$T$-driven conductors}
\label{sec:Scattering}
We consider a mesoscopic two-terminal conductor modeled in the scattering approach as two macroscopic electron reservoirs connected by waveguides to a scattering region \cite{Blanter,Nazarov}. Uncorrelated electrons leave the reservoir and transverse through the scattering region, where they are elastically scattered. Interactions between the electrons and charging effects are disregarded. We denote the hot terminal as $1$ and the cold one as $2$. The Fermi functions $f_\alpha(E)=\{\exp [\beta_\alpha (E-\mu_\alpha)]+1 \}^{-1}$ govern the energy distribution of emanating electrons from terminals $\alpha \in \{1,2 \}$. The parameters $\beta_1 = 1/k_\text{B} T_h$ and $\beta_2 = 1/k_\text{B} T_c$ are determined by the temperatures $T_h$ and $T_c$ of the hot and cold reservoir, respectively. We suppose that there is no applied voltage and set $\mu_1=\mu_2=0$ in the following. The transport is described by the unitary scattering matrix 
\begin{equation}
s(E) = \begin{pmatrix}
s_{11}(E) & s_{12}(E) \\
s_{21}(E) & s_{22}(E)
\end{pmatrix}\,.
\end{equation}
In the submatrix $s_{\alpha \beta}(E)$, the index $\beta$ indicates where the electrons originated and the left index $\alpha$ where the electrons head to. For a given energy, we assume that a finite number of quantum channels contribute, which determines the size of the submatrices. Overall, the transport properties of the mesoscopic conductor are determined by the scattering matrix and the Fermi distributions of the reservoirs. 
\subsection{Noise spectrum and light emission}
The current operator $\hat{I}_\alpha(t)$ describes the electric current in terminal $\alpha \in \{1,2 \}$ at time $t$ \cite{Blanter,Buettiker,Buettiker_1992}. By averaging over the quantum statistical ensemble, the average current $\langle \hat{I}_\alpha(t) \rangle$ is obtained. A positive sign means that the current leaves the reservoir, and vice versa, a negative sign means that it enters. Current fluctuations are characterized by the auto- and cross-correlations of the current operator. We are interested in the nonsymmetrized correlator
\begin{equation}
S_{\alpha \beta}(t,t') = \langle \Delta \hat{I}_\alpha(t) \Delta \hat{I}_\beta(t') \rangle \,, \Delta \hat{I}_\alpha(t)=\hat{I}_\alpha(t)-\langle \hat{I}_\alpha(t) \rangle.
\end{equation} 
In the absence of a time-dependent external field, the correlations depend only on the time difference $t-t'$. As a result, the Fourier transform of the current correlations yields $S_{\alpha \beta}(\omega,\omega')= 2 \pi \delta(\omega+\omega') S_{\alpha \beta}(\omega)$, with $S_{\alpha \beta}(\omega)=2\int e^{i \omega t} \langle \Delta \hat{I}_\alpha(t) \Delta \hat{I}_\beta(0) \rangle dt $ as the noise spectrum. At different times, the current operators do not commute. This results in an asymmetrical noise spectrum. For $\alpha=1$, $\beta=2$, the noise spectrum has the form
\begin{widetext}
\begin{align}
S_{12}(\omega)=\frac{e^2}{ \pi \hbar}  \int dE \bigg [& f_1(E) f^h_1(E+\hbar \omega) \Tr \bigg(s^\dagger_{21}(E+\hbar \omega) s_{21}(E)(s^\dagger_{11}(E)s_{11}(E+\hbar \omega)-1) \bigg ) \nonumber\\
& +f_1(E) f^h_2(E+\hbar \omega)  \Tr \bigg ( s^\dagger_{11}(E) s_{12}(E+\hbar \omega) s^\dagger_{22}(E+\hbar \omega)s_{21}(E) \bigg) \nonumber \\
&+f_2(E) f^h_1(E+\hbar \omega) \Tr \bigg ( s^\dagger_{12}(E)s_{11}(E+\hbar \omega) s^\dagger_{21}(E+\hbar \omega)s_{22}(E) \bigg ) \nonumber\\
&+f_2(E) f^h_2(E+\hbar \omega) \Tr \bigg( s^\dagger_{12}(E)s_{12}(E+\hbar \omega)(s^\dagger_{22}(E+\hbar \omega)s_{22}(E)-1) \bigg) \bigg ],
\end{align}
\end{widetext}
with $e$ the elementary charge, $\hbar$ the reduced Planck constant, $\dagger$ the conjugate transpose, $\Tr(\cdot)$ the trace over the quantum channels and spin, and $f^h_\alpha(E) \equiv 1-f_\alpha(E)$ the distribution of an unoccupied electronic state (hole). In general, $s^\dagger(E)s(E')\neq 1$ for $E\neq E'$ and thus the noise spectrum can be complex. \\
\\
We focus on the total current $\hat{I}(t) = [\hat{I}_{1}(t)-\hat{I}_2(t)]/2$ and investigate the total noise spectrum for this symmetric choice of the currents \cite{Pretre_EtAl_1996,Pedersen_Buettiker_1998,Filipovic_Belzig_2016}
\begin{align}
S(\omega) \coloneqq & 2 \int dt e^{i \omega t} \langle \Delta \hat{I}(t) \Delta \hat{I}(0) \rangle \nonumber \\
=&\sum_{\alpha ,\beta \in \{1,2\}} \int dE \, \gamma_{\alpha \beta}(E,\omega)  f_\alpha(E) f^h_\beta(E+\hbar \omega) ,
\label{eq:noisespec}
\end{align}
with 
\begin{align}
\gamma_{\alpha \beta}(E,\omega) = \frac{e^2}{4 \pi \hbar}  \Tr \bigg(A_{\alpha \beta}(E,\omega) A^\dagger_{\alpha \beta}(E,\omega) \bigg)
\end{align}
and the matrices
\begin{align}
A_{11}=& s^\dagger_{21}(E+\hbar \omega)s_{21}(E)+1-s^\dagger_{11}(E+\hbar \omega)s_{11}(E) \nonumber \\
A_{22}=& s^\dagger_{12}(E+\hbar \omega)s_{12}(E)+1-s^\dagger_{22}(E+\hbar \omega)s_{22}(E) \nonumber \\
A_{12}=& s^\dagger_{21}(E)s_{22}(E+\hbar \omega)-s^\dagger_{11}(E)s_{12}(E+\hbar \omega) \nonumber \\
A_{21}=& s^\dagger_{22}(E)s_{21}(E+\hbar \omega)-s^\dagger_{12}(E)s_{11}(E+\hbar \omega).
\end{align}
Further on, the term total noise spectral density is abbreviated as noise spectrum. The form Eq.~\eqref{eq:noisespec} directly implies that the noise spectrum is positive, since for a complex matrix $A$, the matrix $A^\dagger A$ has only positive eigenvalues and thus a positive trace $\Tr(A^\dagger A) \geq 0$. With the same argument and the monotonic decrease of the Fermi functions, the inequality 
\begin{equation}
S(\omega)\geq S(-\omega) \quad\mathrm{ for }\quad \omega \geq 0
\label{eq:Inequal}
\end{equation}
follows. This result holds irrespective of the applied voltage or temperature difference. In the next paragraph, we introduce the connection to light emission and absorption, where the inequality has an illustrative interpretation.\\
\\
The mesoscopic conductor can act as an antenna, where the current fluctuations couple to an electromagnetic field \cite{Gavish:1,Gavish:2,Lesovik}. We assume a linear coupling between the total current operator $ \hat{I}(t)$ and the electromagnetic vector potential operator. Fermi's golden rule gives the transition rates for absorption and emission of a photon and thus the rate at which energy is transferred \cite{Baym}. To establish a connection, we consider the rewritten noise spectrum
\begin{equation}
S(\omega)=2 \pi \sum_{i,f} p_i | \bra{i} \Delta \hat{I}(0) \ket{f}|^2 \delta (E_i-E_f+\hbar \omega),
\label{eq:Rewnoisespec}
\end{equation}
where $p_i$ is the statistical weight along with $E_i$ the energy of the initial state $\ket{i}$, and $E_f$ the energy of the state $\ket{f}$. The energy transfer rate when emitting one photon from the conductor is proportional to the noise spectrum for $\omega<0$ and vice versa, when absorbing one photon to the noise spectrum for $\omega>0$. Therefore, the noise spectrum at negative frequencies is referred to as the emission spectrum and that of positive frequencies as the absorption spectrum. Utilizing $\Delta \hat{I}(0)$ instead of $\hat{I}(0)$ in expression \eqref{eq:Rewnoisespec} differs by a DC component, which contributes nothing to the energy transfer. The inequality \eqref{eq:Inequal} now states that the energy transfer rate at which the conductor absorbs from the electromagnetic field is always greater than or equal to the rate at which it transfers energy to the field.\\
\\
We follow the interpretation in \cite{Zamoum_Lavagna_2016}. The physical processes involve electrons with energy $E$ and holes with energy $E+\hbar \omega$ which are scattered in the mesoscopic conductor and afterwards recombine to emit or absorb energy $\hbar \omega$. When electrons come from terminal $\alpha$ and holes from terminal $\beta$, then they contribute $\gamma_{\alpha \beta}(E,\omega) f_\alpha(E)f^h_\beta(E+\hbar \omega)$ to the differential energy transfer rate
\begin{align}
dS(E,\omega)=\sum_{\alpha, \beta \in \{1,2\}} \gamma_{\alpha \beta}(E,\omega) f_\alpha(E) f^h_\beta(E+\hbar \omega) dE.
\end{align}  
On average, there are $f_\alpha(E)$ electrons and $f^h_\beta(E+\hbar \omega)$ holes participating in $f_\alpha(E)f^h_\beta(E+\hbar \omega)$ recombination events. This is evident by rewriting $f_\alpha(E)f^h_\beta(E+\hbar \omega)=\langle \hat{n}^e_\alpha(E) \hat{n}^h_\beta(E+\hbar \omega) \rangle$ as the expectation value of the electron number operator $\hat{n}^e_\alpha(E)$ and the hole number operator $\hat{n}^h_\beta(E)$. The functions $\gamma_{\alpha \beta}$ correspond to the rates per recombination event. They depend solely on the scattering amplitudes of the electrons and holes traversing the conductor.

\subsection{Definition of the delta-$T$ noise spectrum}

In the tradition of analyzing noise, two distinct contributions of different origins were identified \cite{Schottky,Nyquist,Johnson,Blanter}. One is the equilibrium noise (Nyquist-Johnson noise) at thermal equilibrium, attributed to the thermal agitation of the charge carriers, and the other is the nonequilibrium partition noise (shot noise), which arises because an incident charge carrier beam is stochastically divided. Hence, it is a consequence of charge quantization. In the next paragraph, we are concerned with the question how can the noise spectrum be partitioned into an equilibrium (thermallike) and a nonequilibrium (delta-$T$) component in the presence of a temperature difference. \\
\\
In the case of zero frequency and an energy-independent scattering matrix, the thermallike noise is given by
\begin{equation}
S^{th}(\omega=0)=S^{th}= 2 G k_\text{B} T_h+ 2 G k_\text{B} T_c,
\end{equation}
with $G=G_0 \sum_n D_n$ the conductance, the conductance quantum $G_0=2e^2/h$, and $D_n$ the transmission probability of the $n$-th eigenchannel \cite{Blanter}. Half of the thermallike noise consists of the Nyquist-Johnson noise at the hot temperature and half at the cold temperature. We interpret that as the thermal agitation of two isolated equilibrium systems and understand the thermallike noise as the average $S^{th}=(S_{T_h}+S_{T_c})/2$ of them, with $S_{T_\alpha}$ the noise at zero frequency and equal temperature $T_\alpha$ in both reservoirs. Following this interpretation, the thermallike noise for finite frequencies is defined as
\begin{equation}
S^{th}(\omega) \coloneqq \frac{S_{T_h}(\omega)+S_{T_c}(\omega)}{2},
\label{eq:Thermalnoise}
\end{equation} 
where $S_{T_\alpha}(\omega)$ is the noise spectrum at equal reservoir temperature $T_\alpha$. From the standpoint of light emission, the thermallike spectrum manifests as the superposition of independent hot and cold emitters. In \cite{Hasegawa} an equivalent definition is used to account for the thermallike part in a system with interacting electrons. After some calculation and using the unitary property of the scattering matrix, the thermallike noise can be written as
\begin{align}
\label{eq:Thermalnoise2}
&S^{th}(\omega)=\int dE \, \gamma^{th}(E,\omega) \sum_{\alpha \in \{1,2\}} \frac{ f_\alpha(E) f^h_\alpha(E+\hbar \omega)}{2} ,
\end{align}
with
\begin{align}
\gamma^{th}(E,\omega)=&\gamma_{11}(E,\omega)+\gamma_{12}(E,\omega)+\gamma_{21}(E,\omega)+\gamma_{22}(E,\omega) \nonumber  \\
=&\frac{e^2}{4 \pi \hbar} \Tr\bigg[ |s_{11}(E)-s_{11}(E+\hbar \omega)|^2 \nonumber \\
&+ |s_{22}(E)-s_{22}(E+\hbar \omega)|^2  \nonumber \\
&+ |s_{12}(E)+s_{12}(E+\hbar \omega)|^2 \nonumber \\
&+ |s_{21}(E)+s_{21}(E+\hbar \omega)|^2  \bigg],
\end{align}
where $|a|^2=a^\dagger a$ is implied in the multi-channel case. At zero frequency, the expression coincides with the thermallike noise defined in \cite{Eriksson} and reduces to
\begin{equation}
S^{th}=\frac{2 e^2}{\pi \hbar} \int dE \sum_n D_n(E) \sum_{\alpha=1,2} f_\alpha(E) f^h_\alpha(E),
\end{equation}
with $D_n(E)$ the energy-dependent transmission probability. \\
\\
Equipped with the definition of the thermallike noise, the nonequilibrium contribution is identified as the excess noise spectrum $S^{\Delta T}(\omega) \coloneqq S(\omega)-S^{th}(\omega)$. This nonequilibrium noise spectrum is referred to as delta-$T$ noise spectrum.  We interpret the delta-$T$ noise spectrum by using the picture of recombining electrons and holes. Electrons are denoted as hot or cold if they are distributed according to the Fermi function $f_1=\langle \hat{n}^e_1 \rangle$ with temperature $T_h$ or according to the Fermi function $f_2=\langle \hat{n}^e_2 \rangle$ with temperature $T_c$, respectively. This designation also applies to holes. The same rates $\gamma_{\alpha \beta}$ are present in the entire noise spectrum \eqref{eq:noisespec} and the thermallike spectrum \eqref{eq:Thermalnoise2}. The difference manifests itself in the number of recombination events. For example, the actual noise spectrum includes the contribution $\gamma_{12} \- \langle \hat{n}^e_1 \hat{n}^h_2 \rangle$, where hot electrons recombine with cold holes. The corresponding term $\gamma_{12}\- \langle(\hat{n}^e_1 \hat{n}^h_1+\hat{n}^e_2 \hat{n}^h_2)/2 \rangle$ in the thermallike noise spectrum involves only recombination events of hot electrons with hot holes and cold electrons with cold holes. This is accounted for by a factor $\gamma_{12}\- \langle(\hat{n}_1^e (\hat{n}^h_2-\hat{n}^h_1)+(\hat{n}^e_1-\hat{n}^e_2) \hat{n}^h_2)/2 \rangle$ in the delta-$T$ noise spectrum, which depends on the excess amount of cold and hot holes $\langle \hat{n}^h_2-\hat{n}^h_1\rangle$ and the excess number of hot and cold electrons $\langle \hat{n}^e_1-\hat{n}^e_2 \rangle$. Another example is the noise contribution $\gamma_{11} \- \langle \hat{n}^e_1 \hat{n}^h_1 \rangle$, which has the counterpart $\gamma_{11} \-\langle(\hat{n}^e_1 \hat{n}^h_1+\hat{n}^e_2 \hat{n}^h_2)/2 \rangle$ in the thermallike noise spectrum. The delta-$T$ contribution assumes the form $\gamma_{11} \langle(\hat{n}^e_1 \hat{n}^h_1-\hat{n}^e_2 \hat{n}^h_2)/2 \rangle$ and depends on the difference between hot and cold recombination events.\\
\\
We distinguish between these two cases and split the delta-$T$ noise spectrum into
\begin{align}
S^{\Delta T}(\omega) \coloneqq & S^{\Delta T}_1(\omega)+S^{\Delta T}_2(\omega) \nonumber \\
=& \int dE \sum_{\alpha \neq \beta \in \{1,2\}} \gamma_{\alpha \beta}(E,\omega) F_{\alpha \beta}(E,\omega)  \nonumber \\
&+ \int dE \left( \gamma_{11}(E,\omega)-\gamma_{22}(E,\omega) \right) F(E,\omega) 
\label{eq:Neqnoise}
\end{align}
with functions
\begin{align}
\Delta f_{\alpha \beta}(E) = & f_\alpha(E)-f_\beta(E) \nonumber \\
F_{\alpha \beta} (E,\omega) = &  \frac{f_\alpha(E) \Delta f_{\alpha \beta}(E+\hbar \omega)+ \Delta f_{\alpha \beta}(E) f^h_\beta(E+\hbar \omega)}{2} \nonumber \\
F(E,\omega) =& \frac{ f_1(E)f^h_1(E+\hbar \omega)-f_2(E)f^h_2(E+\hbar \omega)}{2}.
\end{align} 
The unitary property of the scattering matrix guarantees the symmetry $\Tr A_{12}A_{12}^\dagger=\Tr A_{21}A_{21}^\dagger$. Moreover, $F_{12}(E, \omega)+F_{21}(E, \omega)$ reduces to $\Delta f_{12}(E) \Delta f_{12}(E+\hbar \omega)$ and the first contribution is given by
\begin{equation}
S_1^{\Delta T}(\omega)= \int dE \, \gamma_{12}(E,\omega) \Delta f_{12}(E) \Delta f_{12}(E+\hbar \omega).    
\end{equation}
This contribution stems from recombinations of electrons and holes with different temperatures, e.g., hot electrons with cold holes. It depends only on the excess number of electrons $\langle \hat{n}^e_1-\hat{n}^e_2 \rangle$ and holes $\langle \hat{n}^h_2-\hat{n}^h_1 \rangle$. In the lowest order, the difference $\Delta f_{12}(E)$ is linear in $\Delta T$ and therefore the whole term is quadratic. \\
The thermallike noise spectrum \eqref{eq:Thermalnoise2} includes recombination events of hot electrons with hot holes with a rate $\gamma_{22}$. In contrast to the spectrum \eqref{eq:noisespec}, where this rate is associated with cold electrons recombining with cold holes. A similar situation arises for the rate $\gamma_{11}$, where the thermallike spectrum contains cold recombination events and the actual spectrum contains hot recombination events. These cases are covered by the second contribution $S^{\Delta T}_2(\omega)$. This contribution does not depend on the difference between the Fermi functions but rather on the difference between the number of hot and cold recombination events $\langle \hat{n}^e_1 \hat{n}^h_1-\hat{n}^e_2 \hat{n}^h_2 \rangle$. An expansion in $\Delta T$ results in a linear dependence at the lowest order. For the part that depends on the scattering matrix, we find  
\begin{align}
\Tr \left[{A_{11}A_{11}^\dagger-A_{22}A_{22}^\dagger} \right] =& 2 \Re  \Tr \bigg[ s_{22}(E)s_{22}^\dagger(E+\hbar \omega) \nonumber \\
&+s_{21}(E)s_{21}^\dagger(E+\hbar \omega) \nonumber \\
&-s_{11}(E)s_{11}^\dagger(E+\hbar \omega) \nonumber \\
&-s_{12}(E)s_{12}^\dagger(E+\hbar \omega) \bigg],    
\end{align} 
with $\Re$ denoting the real part. At zero frequency, this expression disappears because of the unitary property of the scattering matrix. The entire spectrum $S_2^{\Delta T}(\omega)$ vanishes in two ways: first, when the scattering properties from both sides are indistinguishable, i.e., $s_{11}(E)=s_{22}(E), s_{12}(E) = s_{21}(E)$, or second, the reservoirs assume the same temperature.  \\ 
At zero frequency, the delta-$T$ noise decreases to
\begin{equation}
S^{\Delta T}=\frac{2 e^2}{\pi \hbar} \int dE \sum_n D_n(E) [1-D_n(E)] [\Delta f_{12}(E)]^2
\end{equation}
and matches with the definition given in \cite{Eriksson}.
\section{Energy-independent scattering}
\label{sec:IndepEnergy}
\begin{figure}
\includegraphics[scale=1]{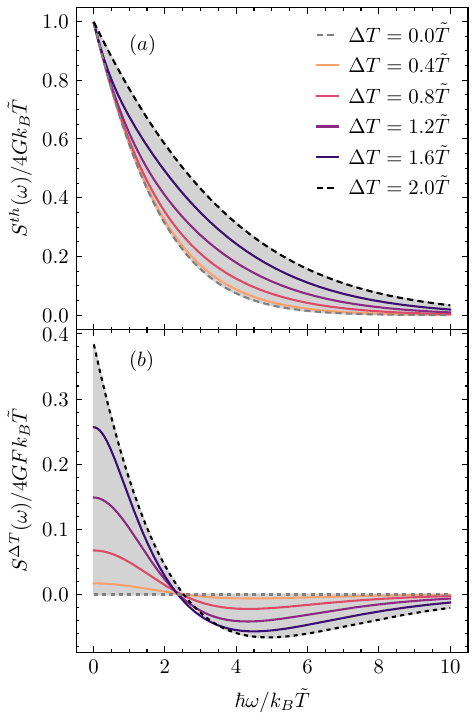}
\caption{Emission spectra of $(a)$ the thermallike noise \eqref{eq:IndepThermalnoise} and $(b)$ the delta-$T$ noise \eqref{eq:IndepDTnoise} from $\Delta T=0$ to $2 \tilde{T}=T_h+T_c$. The absorption spectrum of $(a)$ results from $S^{th}(\omega)-S^{th}(-\omega)=4G \hbar \omega$ and of $(b)$ equals the emission spectrum, because of $S^{\Delta T}(\omega)=S^{\Delta T}(-\omega)$. The thermallike spectrum depends on the temperature difference $\Delta T$ and all curves lie between the extreme points $\Delta T=0$ and $\Delta T=2 \tilde{T}$ in the grey area. The delta-$T$ noise changes its sign from positive to negative at a certain frequency (compare to Fig. \ref{fig:Integ}), which depends on the temperature difference. The curves are not enveloped by the extreme points. The overall noise $S^{th}+S^{\Delta T}\geq 0$ does not change its sign.}
\label{fig:noise}
\end{figure}

The model is specified further by assuming an energy-independent scattering matrix $s(E) \approx s$. From the well-known equilibrium spectrum \cite{Lesovik,Blanter}, we obtain  
\begin{equation}
S^{th}(\omega)=G \hbar \omega \left[2+\coth \left(\frac{\hbar \omega}{2 k_\text{B} T_h} \right)+ \coth \left(\frac{\hbar \omega}{2 k_\text{B} T_c} \right) \right].
\label{eq:IndepThermalnoise}
\end{equation}
The thermal spectra $S_{T_h}(\omega),S_{T_c}(\omega)$ are not additive, i.e., $S_{T_h}(\omega)+S_{T_c}(\omega)$ cannot be written as a thermal noise spectrum with adjusted temperature. A consequence of this is that at a fixed average temperature $\tilde{T}=(T_h+T_c)/2$, the thermallike spectrum still changes for different temperature differences $\Delta T=T_h-T_c$. Nevertheless, for convenience, the average temperature and the temperature difference are used hereafter.\\
\\
The difference between the absorption and emission spectrum obeys $S^{th}(\omega)-S^{th}(-\omega)=4G \hbar \omega, \omega>0$. It is sufficient to investigate the emission spectrum since emission and absorption differ only by the zero-point fluctuations. Figure \ref{fig:noise}$(a)$ depicts the thermallike spectrum for various $\Delta T$. 
The curves vary continuously from $\Delta T=0$ to the maximum $\Delta T=2 \tilde{T}$, where the curve for $\Delta T=0$ sets a lower bound and for $\Delta T=2 \tilde{T}$ an upper bound. \\
\\
The examination of the delta-$T$ noise spectrum reveals that $S^{\Delta T}_2(\omega)$ vanishes for energy-independent scattering, and the delta-$T$ component reduces considerably to
\begin{equation}
S^{\Delta T}(\omega)= 2GF \int dE \Delta f_{12}(E) \Delta f_{12}(E+\hbar \omega),
\label{eq:IndepDTnoise}
\end{equation}
where $F=\sum_n D_n (1-D_n)/\sum_n D_n$ represents the Fano factor. The Fano factor indicates for the property of a partition noise and is a shared property with the shot noise at zero frequency \cite{Blanter}. If the barrier is completely transparent or reflective, $D_n=1$ or $D_n=0$, $\forall n$ , then the Fano factor vanishes and so does the delta-$T$ noise spectrum, although a temperature difference might be present. The system consists then of isolated hot and cold electron transport, each in thermal equilibrium, which is why we call the term thermallike.\\
At the maximal temperature difference $\Delta T=2 \tilde{T}$, it is given as 
\begin{align}
\frac{ S^{2 \tilde{T}}(\omega)}{GF}=& 8 k_\text{B} \tilde{T}  \ln \left[2 \cosh \left(\frac{\hbar \omega}{4 k_\text{B} \tilde{T}} \right) \right] \nonumber \\
	& - |\hbar \omega|- \hbar \omega \coth \left(\frac{\hbar \omega}{4 k_\text{B} \tilde{T}} \right).
\end{align}
In the limit $\omega \rightarrow 0$, we get back the factor $2 \ln 2$ as in \cite{Larocque}. The delta-$T$ emission and absorption spectrum for different $\Delta T$ is shown in Fig. \ref{fig:noise}(b). The delta-$T$ noise spectrum starts at a maximum, then decreases and intersects the $\omega$-axis at a certain point. After that, the course reaches a minimum and converges back to zero. Above a certain point, the delta-$T$ noise spectrum is negative and decreases the thermallike noise. 
\begin{figure}
\includegraphics[scale=1]{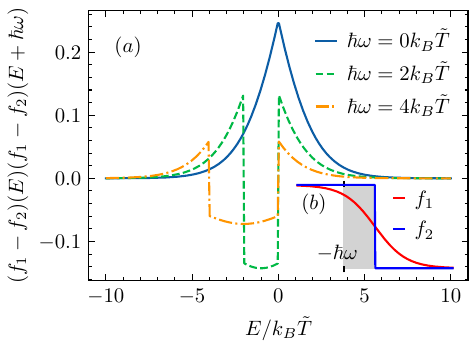}
\caption{Illustration $(a)$ shows the excess number of recombination events due to recombinations between hot and cold particles, i.e., the integrand in \eqref{eq:IndepDTnoise}. At a given frequency (here e.g., $\hbar \omega=0,2,4 k_\text{B} \tilde{T}$), one of the differences $\Delta f_{12}$ is shifted by $\hbar \omega$ in relation to the other. In the interval $(-\hbar \omega,0)$, the contribution is negative and outside it is positive. Inset $(b)$ depicts the hot and cold Fermi functions drawn over energy. The gray-shaded area denotes energies where the sign of $f_1(E)-f_2(E)$ and the shifted $f_1(E+\hbar \omega)-f_2(E+\hbar \omega)$ differ. The terminals are assumed to be at temperatures $T_c=0, T_h=2 \tilde{T}$.}
\label{fig:Integ}
\end{figure}
Negative delta-$T$ noise (at zero frequency) has been reported for transport in a fractional quantum Hall bar \cite{Rech}. The decrease in fluctuations is there attributed to interactions and related to the tunneling of quasiparticles. In our case, interactions are not considered by the model. The combination of temperature broadening of the Fermi function and the frequency shift between distributions of the occupied and free states plays the crucial role here. A comparison between the cold and hot distribution shows (see for an example Fig. \ref{fig:Integ}$(b)$), that the hot reservoir has fewer occupied states for $E<0$, more for $E>0$ and coincides for $E=0$. This leads to a negative sign of $f_1(E)-f_2(E)$ for $E<0$ and a positive sign for $E>0$. The integrand $(f_1(E)-f_2(E))(f_1(E+\hbar \omega)-f_2(E+\hbar \omega))$, i.e., the excess number of recombination events, has a negative sign in the interval $E \in (-\hbar \omega,0)$. In the complement interval, the signs are equal and the integrand is positive. The delta-$T$ noise spectrum turns negative when the area under the negative part of the integrand exceeds the area under the positive part. Figure \ref{fig:Integ}$(a)$ depicts the excess number of recombination events for different shifts in the extreme case $T_c=0$ and $T_h=2 \tilde{T}$.

\section{Resonant level}
\label{sec:Resonance}
\begin{figure}
\includegraphics[scale=1]{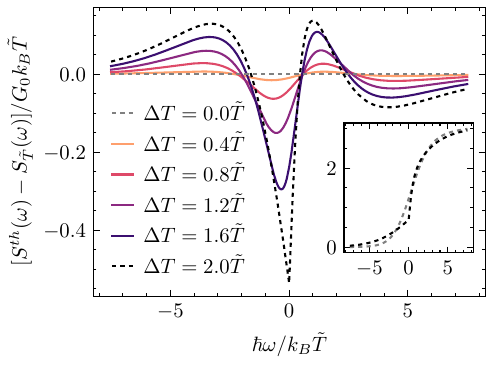}
\caption{The deviation of the thermallike noise spectrum \eqref{eq:ResThermalnoise} from the thermal spectrum at average temperature $S_{\tilde{T}}(\omega)$ is shown for various temperature differences. In the inset, we depict the thermallike noise spectrum at the extremes $\Delta T=0$, $\Delta T=2 \tilde{T}$ in order to give an idea of its shape and size. The units are the same as in the whole figure. The resonant level is located at $\epsilon_0=0$, i.e., at the terminal's chemical potential, and has the widths $\Gamma_1=(2/3)k_\text{B} \tilde{T}$, $\Gamma_2=(1/3) k_\text{B} \tilde{T}$.}
\label{fig:ResThermalnoise}
\end{figure}

This paragraph sheds light on the influence of energy-dependent scattering, using a resonant level as an example. We assume a resonance energy of $\epsilon_0$ and a lifetime of $\hbar/\Gamma$. For simplicity, we only consider one open quantum channel. The scattering matrix can be modeled by the Breit-Wigner formula \cite{Breit,Buettiker} 
\begin{equation}
s_{\alpha \beta}(E)=\delta_{\alpha \beta}-\frac{ i \sqrt{\Gamma_\alpha \Gamma_\beta}}{E-\epsilon_0+i \Gamma/2}, \, \alpha,\beta \in \{1,2\}
\end{equation} 
where $\Gamma_1,\Gamma_2$ are the partial widths, $\Gamma=\Gamma_1+\Gamma_2$ the total width, and $\delta_{\alpha \beta}$ the Kronecker delta. Inserting this scattering matrix into the thermallike noise spectrum \eqref{eq:Thermalnoise2} results in
\begin{align}
&S^{th}(\omega)=\frac{G_0}{4} \int dE \sum_{\alpha=1,2} f_\alpha(E) f^h_\alpha(E+\hbar \omega) D(E+\hbar \omega) \nonumber \\
&\times D(E) \hspace{-3pt}  \left[\frac{(\hbar \omega)^2}{\Gamma^2_1}+\frac{(\hbar \omega)^2}{\Gamma^2_2}+8\frac{ (E+\frac{\hbar \omega}{2}-\epsilon_0)^2+\frac{\Gamma^2}{4}}{\Gamma_1 \Gamma_2} \right],
\label{eq:ResThermalnoise}
\end{align} 
where 
\begin{equation}
D(E)=\frac{\Gamma_1 \Gamma_2}{(E-\epsilon_0)^2+\Gamma^2/4}
\end{equation}
represents the transmission probability through the resonance. Figure \ref{fig:ResThermalnoise} depicts the deviation of thermallike spectrum from the thermal spectrum at average temperature. We fix the average temperature and consider different temperature differences for an asymmetric resonance $\Gamma_1=(2/3) k_\text{B} \tilde{T}, \Gamma_2=(1/3) k_\text{B} \tilde{T}$. The curves for $\Delta T \neq 0$ intersect with the curve for $\Delta T=0$ at three points and have a smaller value in two intervals. 
\begin{figure}
\includegraphics[scale=1]{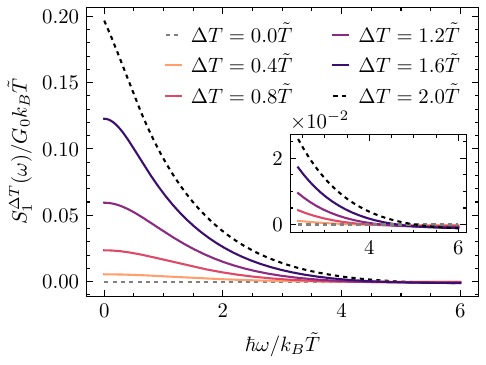}
\caption{Spectra of the contribution \eqref{fig:ResdT1} at different $\Delta T$. Emission and absorption spectra are identical because this delta-$T$ contribution is symmetric in frequency. The resonance is assumed to be at $\epsilon_0=0$ and has the widths $\Gamma_1=(2/3)k_\text{B} \tilde{T}$, $\Gamma_2=(1/3) k_\text{B} \tilde{T}$. The inset shows the spectra at the transition from positive to negative. This negative part is almost suppressed but still present.}
\label{fig:ResdT1}
\end{figure}
\begin{figure}
\includegraphics[scale=1]{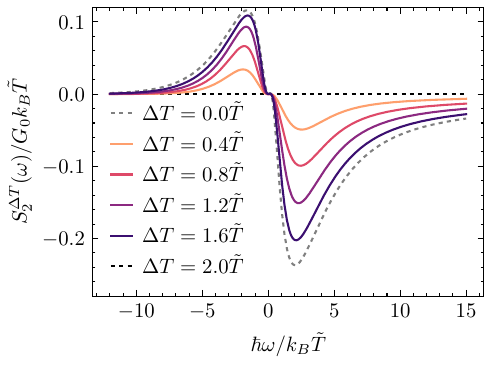}
\caption{Spectra of the second component \eqref{eq:ResdT2} at different temperature differences $\Delta T$. The partial widths are $\Gamma_1=(2/3)k_\text{B} \tilde{T}$, $\Gamma_2=(1/3) k_\text{B} \tilde{T}$, and the resonance energy is $\epsilon_0=0$. In the emission spectrum, we see an enhancement, and in the absorption spectrum, a reduction.}
\label{fig:ResdT2}
\end{figure}\\
\\
The resonance provides an example where both contributions in the delta-$T$ noise spectrum are relevant. We write the first contribution in the form
\begin{align}
&S^{\Delta T}_1(\omega)=2G_0 \int dE  \Delta f_{12}(E) \Delta f_{12}(E+\hbar \omega)    \nonumber \\
&\times D(E) D(E+\hbar \omega) \left[\frac{(E+\frac{\hbar \omega}{2}-\epsilon_0)^2+\frac{\Gamma^2}{4}}{\Gamma_1 \Gamma_2}-1 \right]  .
\label{eq:ResdT1}
\end{align}
Figure \ref{fig:ResdT1} depicts this contribution for several temperature differences. In the integral \eqref{eq:ResdT1}, the Fermi distributions enter in the same way as for the energy-independent scattering. Temperature broadening in connection with the frequency shift leads again to a negative integrand in the interval $(-\hbar \omega,0)$. The difference is that the resonance introduces an additional weight, which emphasizes, for the chosen parameters, the positive areas of Fig. \ref{fig:Integ}(a) and leads to a suppression of the negative contribution. \\
\\
If the resonance is asymmetric $\Gamma_1 \neq \Gamma_2$, then we obtain a nonvanishing second component
\begin{align}
S^{\Delta T}_2(\omega) =& \frac{G_0 }{4} (\hbar \omega)^2  \frac{\Gamma_1^2 -\Gamma_2^2}{\Gamma_1^2 \Gamma_2^2} \nonumber \\
&\times \int dE  D(E) D(E+\hbar \omega) F(E,\omega),
\label{eq:ResdT2}
\end{align}
which is not symmetric in frequency. As a consequence, the emission spectrum differs from the absorption spectrum. The spectrum exhibits a peak at negative frequencies and thus an enhancement in the emission. For positive frequencies, a dip occurs, which results in a reduction of the absorption. It is the other way round if $\Gamma_1<\Gamma_2$ is used instead of $\Gamma_2<\Gamma_1$. First and second component are of similar size. In Figure \ref{fig:ResdT2}, the second component $S^{\Delta T}_2(\omega)$ is shown for different temperature differences. 

\section{Conclusion}
\label{sec:conclusions}
We have investigated the nonsymmetrized noise spectral density in mesoscopic conductors subjected to a temperature difference. The thermallike spectrum was identified as the average thermal noise spectrum $(S_{T_h}(\omega)+S_{T_c}(\omega))/2$ and, consequently, the delta-$T$ noise spectrum as the excess spectrum. We further decomposed the delta-$T$ noise spectrum into two contributions, one depending on the frequency-shifted differences of Fermi functions and the other on the difference of the combined Fermi functions. In the case of energy-independent scattering, only the first contribution survived, and the spectrum was proportional to the Fano factor. We have discovered a partially negative delta-$T$ noise spectrum that is positive at low frequencies and becomes negative above a certain frequency. A similar behavior, but with a suppressed negative part, was obtained for a resonant level with the selected parameters. In addition, the second component occurred for an asymmetric resonance.

\section*{Acknowledgment}

This research was supported by the Deutsche Forschungsgemeinschaft (DFG; German Research Foundation) via SFB 1432 (Project \mbox{No.} 425217212).

\bibliography{Bibliothek}

\end{document}